\documentclass[12pt]{iopart}
\usepackage{listings}
\usepackage{graphicx}
\usepackage{epstopdf}
\usepackage{amsmath}
\usepackage{multirow}
\usepackage{array}
\usepackage{lineno}

\newcommand{\total}{\textrm{total}}
\newcommand{\IAD}{\textrm{IAD}}
\newcommand{\Mie}{\textrm{Mie}}
\newcommand{\MieRayleigh}{\textrm{Mie+Ray}}
\newcommand{\Ray}{\textrm{Ray}}
\newcommand{\RevIAD}{\textrm{RevIAD}}
\newcommand{\RevIADMie}{\textrm{RevIAD}_{\textrm{Mie}}}
\newcommand{\RevIADMieRay}{\textrm{RevIAD}_{\textrm{Mie+Ray}}}
\newcommand{\msim}{\textrm{sim}}
\newcommand{\simMC}{\textrm{sim.MC}}
\newcommand{\simIAD}{\textrm{sim.IAD}}
\newcommand{\simIADMie}{\textrm{sim.IAD}_{\textrm{Mie}}}
\newcommand{\simIADMieRay}{\textrm{sim.IAD}_{\textrm{Mie+Ray}}}
\newcommand{\simRevIAD}{\textrm{sim.RevIAD}}

\newcommand{\Hb}{\textrm{Hb}}
\newcommand{\HbOO}{\textrm{HbO}_{2}}
\newcommand{\Water}{\textrm{Water}}
\newcommand{\Lipid}{\textrm{Lipid}}
\newcommand{\MetHb}{\textrm{MetHb}}
\newcommand{\IL}{\textrm{IL}}
\newcommand{\calculated}{\textrm{cal}}
\newcommand{\expected}{\textrm{exp}}

\usepackage{hyperref}
\usepackage[capitalise,noabbrev]{cleveref}

\begin{document}

\title[Revised IAD for characterisation of optical properties of small sample]{Methodology development and evaluation of optical properties characterisation of small size tissue samples}

\author{Yijing Xie$^1$, Jonathan Shapey$^{1,2}$, Eli Nabavi$^{1}$, Peichao Li$^{1}$, Anisha Bahl$^{1}$, Michael Ebner$^{1}$,
and Tom Vercauteren$^1$}

\address{$^1$ School of Biomedical Engineering \& Imaging Sciences, King's College London, London SE1 7EH, UK}
\address{$^2$ Department for Neurosurgery, King's College Hospital, London UK}
\ead{yijing.xie@kcl.ac.uk}

\begin{abstract}
Integrating sphere (IS) techniques combined with an inverse adding doubling (IAD) algorithm have been widely used for determination of optical properties of \textit{ex vivo} tissues. Semi-infinite samples are required in such cases. The aim of this study is to develop a methodology for calculating the optical absorption and reduced scattering of biological tissues of small size from IS measurements at 400-1800 nm. We propose a two-stage IAD algorithm to mitigate profound cross-talk effects in the estimation of the $\mu'_s$ in the case of very high $\mu_a$. We developed a small sample adaptor kit to allow IS measurements of samples with small sizes using a commercial spectrophotometer. Results showed that: The two-stage IAD substantially eliminated the cross-talks in the $\mu'_s$ spectra, thus rectifying $\mu_a$ accordingly; and the small sized sample measurements led to systematically overestimated $\mu_a$ values while the spectrum shape well preserved as compared to the normal port size measurements.
\end{abstract}

\vspace{2pc}

\maketitle
\section{Introduction}
Optical imaging modalities have become valuable tools in clinical applications, providing real-time structural, molecular and functional information of tissue for diagnosis, and for guiding surgical or therapeutic interventions \cite{Xie2010FiberImaging,Ebner2021IntraoperativeTranslation,Shapey2019IntraoperativeStudies,Stummer2014,Zharkikh2020BiophotonicsMellitus,Desroches2018ABiopsy,Clancy2021IntraoperativeImaging,Sibai2019QuantitativeResection}. The design and development of optical imaging systems requires a thorough understanding of the optical properties of tissue which are typically characterized by the absorption coefficient $\mu_a$ and reduced scattering coefficient $\mu'_s$.

Tissue optical properties can be estimated using direct or indirect methods \cite{Tuchin2015TissueDiagnosis,Bashkatov2016TissueProperties}. Direct methods are based on fundamental tissue-light interaction rules, such as the Beer-Lambert Law; Direct methods often involve rigorous experimental conditions, such as very thin tissue samples (200 - 500 $\mu$m) to allow only single scattering events.
Indirect methods obtain the solution of the inverse problem using a theoretical model that approximates light propagation in a turbid medium. Indirect methods may be further divided into non-iterative and iterative models. The former are typically based on the two flux Kubelka–Munk model or multi-flux models, that use equations where the optical properties are defined through parameters directly related to the quantities being measured~\cite{Bashkatov2016TissueProperties}.
In indirect iterative methods, the optical properties are implicitly defined through measured parameters; a numerical solution to the model is obtained by iteratively making guesses at the solution, testing whether the problem is solved well enough, and stop accordingly.
Typical approaches include inverse adding doubling (IAD) and inverse Monte Carlo simulation (IMC)\cite{Bashkatov2016TissueProperties}.

Among the aforementioned methods, the IAD algorithm together with measurements stemming from an integrating sphere (IS) technique is one of the most widely used approaches to derive optical properties of \textit{ex vivo} biological tissue samples.
The open-source IAD software \cite{Prahl1993DeterminingMethod} may be conveniently used with IS measurements where the geometrical parameters such as dimension and port size of an integrating sphere, number of measurements, size of the incident light beam and as well as the refractive index of the given biological sample have been considered.
Moreover, it also includes a Monte Carlo subroutine that calculates the loss of the light scattered through the transverse boundary of the sample so as to eliminate the light loss effect on IAD~\cite{Prahl1993DeterminingMethod}.

To date, the majority of the researches concerning tissue optics have employed light in the visible and near-infrared regions of the spectrum (VIS-NIR 400 nm to 1100 nm), as summarised in the work by Madsen and Wilson \cite{Madsen2013OpticalTherapy}, and the work by Tuchin \cite{Tuchin2015TissueDiagnosis}.
Researches focus on characterising the concentration of oxygenated ($\Hb$) and de-oxygenated haemoglobin ($\HbOO$) in a tissue would deploy light in the visible range to be able to detect the primary absorption peaks of $\Hb$ and $\HbOO$ \cite{Rejmstad2017OxygenTrajectories,Crane2003EvidenceAnimal,Giannoni2018HyperspectralDevelopments}.
Light at the long wavelength of 900 to 1100 nm would be utilised in studies focus on the investigation of tissue's water and lipid concentrations \cite{Nachabe2010EstimationNm,Wisotzky2019Determination800nm}.
Furthermore, the tissue samples previously studied were either in large size that were able to cover a standard sphere port or alternatively patched together \cite{Mesradi2013OpticalRat}.

The aim of this study is to develop a methodology for calculating optical absorption ($\mu_a$) and reduced scattering ($\mu'_s$) of small sized (approx. 6 x 10 mm) biological tissues from reflectance and transmittance measurements acquired over an extended spectral range spanning from 400 to 1800 nm.
We developed a small sample adaptor kit to allow integrating sphere measurements of samples with small sizes using a commercial spectrophotometer.
A dual beam single integrating sphere technique was used to measure the total transmittance ($T_{\total}$) and total reflectance ($R_{\total}$) of samples, and inverse adding doubling (IAD) was used to calculate the absorption coefficient and reduced scattering coefficient from the measurements.
We observed that the standard IAD algorithm would produce profound cross-talk effect on the calculated reduced scattering spectrum at wavelength regions where the corresponding absorption coefficient is 
very
high due to high haemoglobin or water content in the tissue sample.
We propose a two-stage IAD algorithm incorporating Mie and Rayleigh Scattering models as input constrains to mitigate this profound cross-talk effect and to provide an accurate estimation of the reduce scattering spectrum of tissue.
We evaluated the two-stage IAD with data generated using Monte Carlo simulation and with rigorous experimental data obtained from controlled phantom experiments.
We then investigated the accuracy of the proposed method using our small sample adaptor kit and condensed incident light beam by comparing results to those obtained with standard IS measurements and IAD with larger samples.

\section{Materials \& methods}
\subsection{Integrating sphere measurement}
\begin{figure}
	\centering
	\includegraphics[scale=0.15]{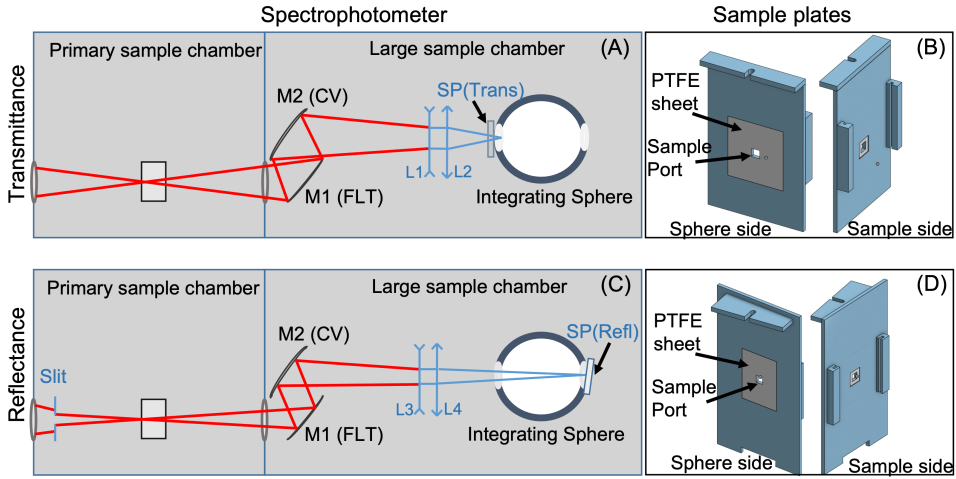}
	\caption{Illustration of the integrating sphere set-up, and our proposed small sample adaptor kit (annotated in blue color), and CAD design of sample plates for transmittance (A, B) and reflectance (C, D) measurements. The small sample adaptor kit includes a lens assembly (L1, L2; L3, L4 ) and a sample plate (SP, position in the set-up shown in (A) and (C)) for adjusting sample aperture size and holding sample slide. CV: Concave, FLT: Flat.}
	\label{fig:1}
\end{figure}

In this study, we used a commercial dual beam high performance spectrophotometer system (Lambda 750s, Perkin Elmer) equipped with a 100 mm single integrating sphere module (L6020371, Perkin Elmer) to obtain transmittance and reflectance measurements.
The system is supplied with a photomultiplier tube detector (PMT, 200 - 860 nm) and a Indium gallium arsenide detector (InGaAs, 860 nm - 2500 nm), covering a wide wavelength range from 200 to 2500 nm.
The system's performance in terms of wavelength calibration and measurement accuracy was validated by the manufacturer and has been regularly serviced according to protocol.
It has been used in research and industrial labs as a standard tool for precise analysis of materials in various applications \cite{Jernshj2009AnalysisSetup,Gaigalas2013MeasurementNm,Jang2015SolidSystems,Dantuma2019Semi-anthropomorphicPhantom,Rehman2020BiomedicalReview,Kanniyappan2020PerformanceImaging}.
A schematic illustration of the original spectrophotometer setup and the modified set up with small sample adaptor kits presenting is shown in \cref{fig:1} (A, C). The original sample port size at the transmittance port is 11 mm (w) $\textrm{x}$ 24 mm (h), and at the reflectance port it is 17 mm (w) $\textrm{x}$ 22 mm (h); the incident light beam size is measured around 3 mm (w) $\textrm{x}$ 15 mm (h) at the transmittance port, and 6.5 mm (w) $\textrm{x}$ 13.5 mm (h) at the reflectance port. The basic operational principle of performing reflectance and transmittance measurements using a spectrophotometer with an integrating sphere, respectively, are also sketched in \cref{fig:1}(A, C).
All measurements were carried out relative to a diffuse optical standard, Spectralon\textsuperscript{\textregistered} (SRS-99-020, Labsphere), which is a white Lambertian material with diffuse reflectance values between 95–99\% in the wavelength region from 250 nm to 2500 nm. With the aforementioned integrating sphere configuration, total transmittance and total reflectance are measured for all experiments conducted in this study and are denoted as \(T_{\total}\) and \(R_{\total}\), respectively, unless otherwise stated.

\subsection{Small sample kit }
The original setup of the spectrophotometer can be used to obtain measurements of various solid materials.
The size of the sample should be such that the distance (\(h\)) from the edge of the incident beam on the sample to the edge of the integrating sphere port is made as large as practicable.
This is to prevent light loss from the lateral sides of the sample. It has been demonstrated that when \(h\) is larger than five times of \(1/(\mu_a + \mu_s')\) the light loss from the lateral sides could be considered negligible \cite{Mesradi2013OpticalRat}.
In order to perform transmittance and reflectance measurements on specimens that are only available in small size (smaller than the original spectrophotometer port size and the beam size on both ports), we designed and developed a small sample kit to adapt the original spectrophotometer setup.
Our small sample kit includes: 1) an adjustable slit to reduce the beam dimension at the beam exit port in the primary sample chamber (shown in \cref{fig:1}(C)); 2) lens sets to focus the incident beam at the transmittance and reflectance ports, respectively; and 3) a sample holder to hold the sample slide and to create a mask on the original port with reduced port size (shown in \cref{fig:1} (B, D)).
The side facing the integrating sphere of the small sample holder consists of a patch of polytetrafluoroethylen (PTFE) sheet (99\% reflectance, Zenith Lite, Pro-Lite Technology, UK ) that covers the original port of the integrating sphere.
In the centre of the PTFE sheet we designed a square port defining the new sample port size.
The PTFE material matches the material of the integrating sphere inner wall material, namely Spectralon\textsuperscript{\textregistered}.
We developed two sets of the small sample kit with configurations listed in \cref{tab.1}. 

\begin{table}
\caption{\label{tab.1}Parameters of our small sample kit}
\footnotesize
\begin{tabular}{@{}lllll}
\br
Reduced port size&Measurement type&Lenses&Slit width&Beam size at the port\\
\mr
3 mm &Transmittance&L1 -150, L2 60&5 mm&1 x 1.5 mm\\
3 mm &Reflectance&L3 -150, L4 200&3 mm&2 x 2 mm\\
5 mm &Transmittance&L1 -150, L2 60&5 mm&1 x 1.5 mm\\
5 mm &Reflectance&L3 -150, L4 200&3 mm&2 x 2 mm\\
\br
\end{tabular}\\
L1: LC1611 Thorlabs; L2: LA1401 Thorlabs; L3: LC1611 Thorlabs; L4: LA1979 Thorlabs
\end{table}

\subsection{Two-stage inverse adding doubling}
\label{sec.reviad}
\begin{figure}
	\centering
	\includegraphics[scale=0.35]{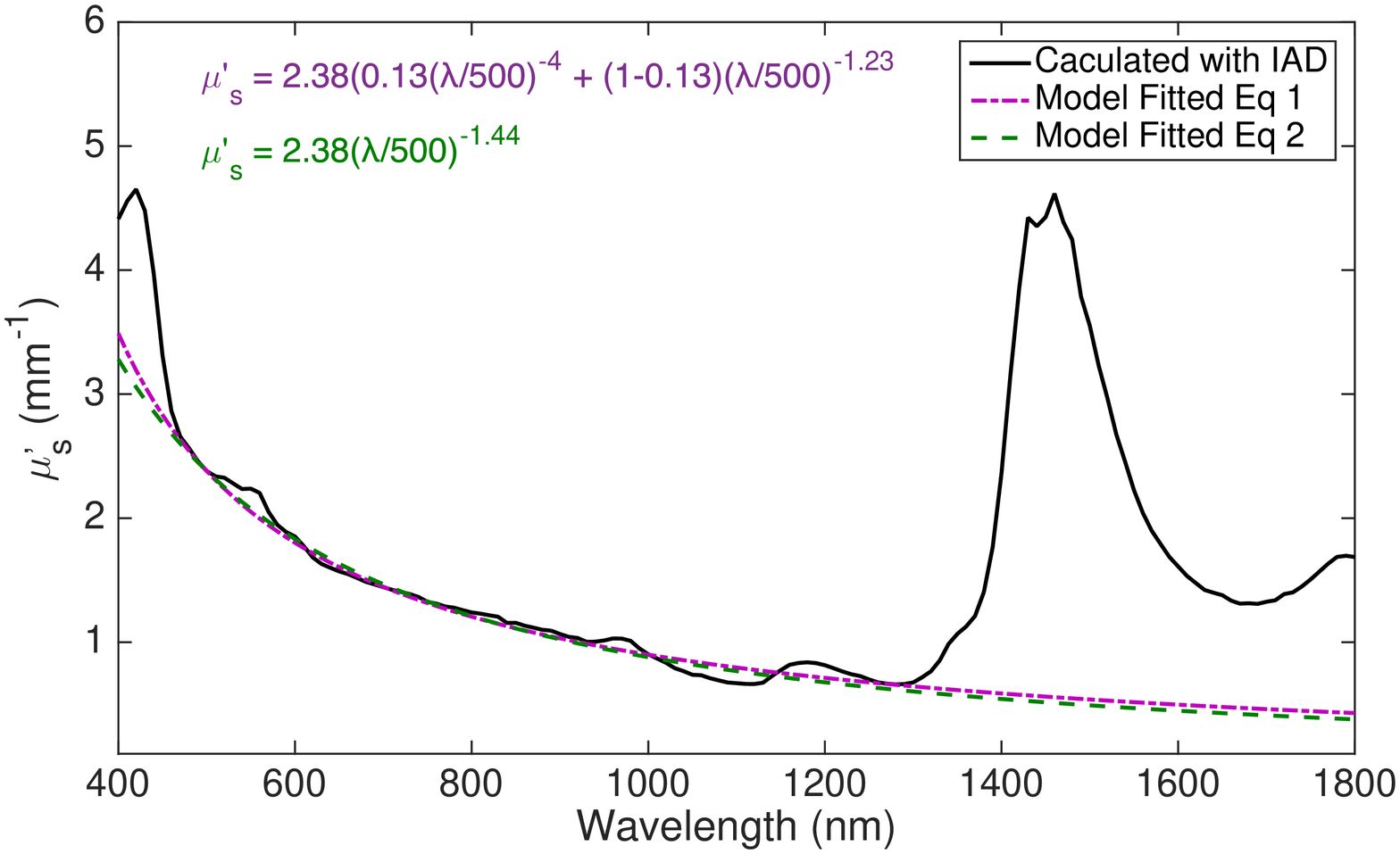}
	\caption{A representative sample of reduced scattering that was calculated from the measured total transmittance and total reflectance using the IAD algorithm. There are noticeable discrepancies at 420 nm, 560 nm and 1450 nm, between the calculated $\mu'_s$ and the ones fitted by Mie / Rayleigh theory. }
	\label{fig:2}
\end{figure}

To derive optical absorption and reduced scattering spectra from transmittance and reflectance spectra, we developed a two-stage IAD algorithm based on the open-source IAD algorithm developed by Prahl et al. \cite{Prahl1993DeterminingMethod}\footnote{Available from \url{https://github.com/scottprahl/iad}}.
In general, inverse adding doubling is a numerical approach to solve the radiative transport equation (RTE) providing analytical calculations of the optical properties of a sample slab from reflectance and transmittance measurements.
The algorithm first assumes a thin slab with known optical properties and calculates the transmittance and reflectance at the slab surface, doubles them as adding another thin layer slab of the same material iteratively until the measured values of transmittance and reflectance are matched with those corresponding to the estimated optical properties.
The algorithm can be conveniently used with integrating sphere measurements where the geometrical parameters such as the port size of the integrating sphere, number of measurements, dimensions of the sphere, size of the incident light beam and the refractive index of the given biological sample have been considered and accounted in the algorithm.
However, a profound cross-talk effect occurs in the IAD calculated reduced scattering at the wavelength regions where the sample's absorption is very high.
This is notably the case at 420 nm where the haemoglobin absorption peak is, and at 1450 nm where the predominant water absorption peak is.
This was observed in our tissue experiments, as shown in \cref{fig:2} and in \cite{Shapey2020ExTissue,Shapey2021OpticalWindow}, and as well as reported and investigated in the work by Gebhart et al. \cite{Gebhart2006InAdding-doubling}. 

\begin{figure}
	\centering
	\includegraphics[scale=0.13]{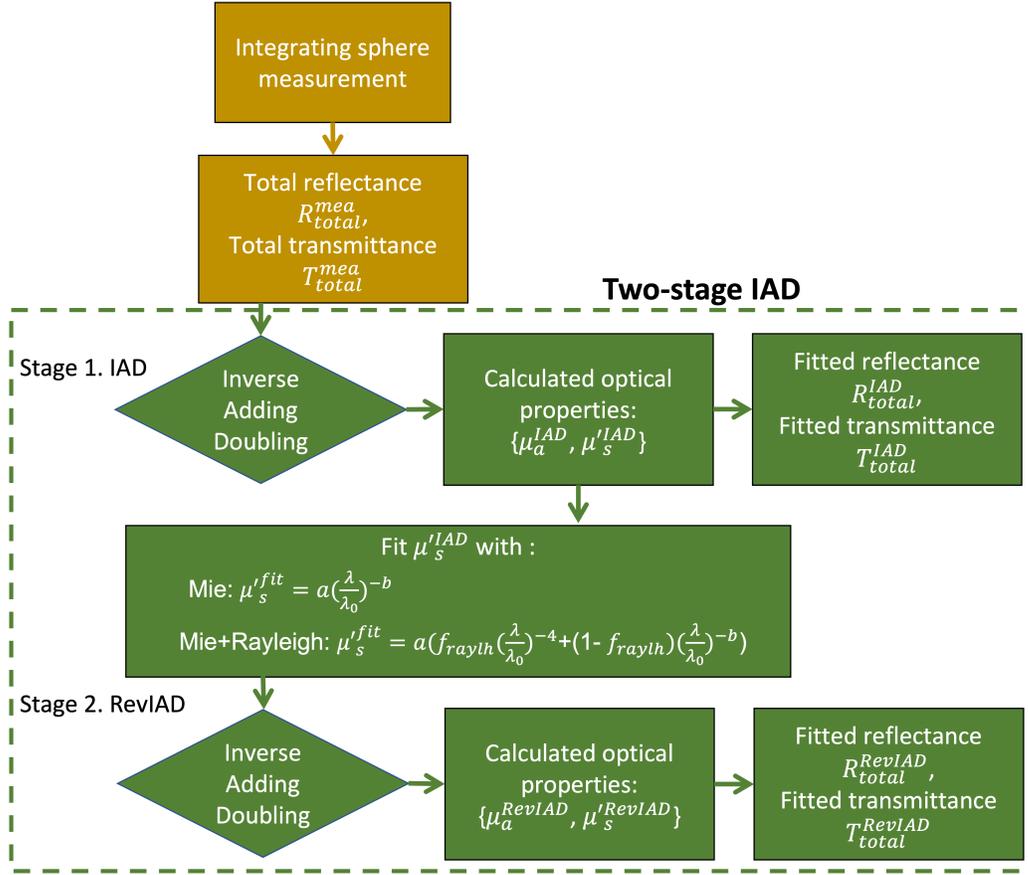}
	\caption{The pipeline of our proposed two-stage IAD algorithm.}
	\label{fig:3}
\end{figure}

To mitigate the inter-parameter cross-talk effect, we propose a two-stage inverse adding doubling process as illustrated in \cref{fig:3}. During the \textbf{first-stage IAD}, for each wavelength $\lambda$, we calculate intermediate values $\mu^{\IAD}_a(\lambda)$ and $\mu'^{\IAD}_s(\lambda)$.
We then used non-linear least squares \cref{eq.6}
\cite{Seber1989NonlinearRegression} to fit the $\mu'^{\IAD}_s(\cdot)$ spectrum with a parametric model.
\begin{equation}
\hat{\beta} \in \textrm{argmin}_{\beta}\sum_{i=1}^{n}[y_i - f(x_i,\beta)]^2 
\label{eq.6}
\end{equation}
In its simplest form, the fitted model is a power law according to the Mie theory:
\begin{equation}
\mu'^{\Mie}_s (\lambda) = a_0\left(\frac{\lambda}{\lambda_0}\right)^{b_{\Mie}}
\label{eq.1}
\end{equation}
where $a_0 = \mu'^{\IAD}_s (\lambda_0)$, $\lambda_0$ is a reference wavelength, $b_{\Mie}$ is the power law exponent (scattering power) related to the Mie scattering.
To account for the contribution of both Mie and Rayleigh scattering, a combined power law may also be fitted:
\begin{equation}
\mu'^{\MieRayleigh}_s (\lambda) = a_0\Big(f_{\Ray}\left(\frac{\lambda}{\lambda_0}\right)^{-4} + (1-f_{\Ray})\left(\frac{\lambda}{\lambda_0}\right)^{b_{\Mie}}\Big)
\label{eq.2}
\end{equation}
The Rayleigh scattering is expressed as $f_{\Ray}(\frac{\lambda}{\lambda_0})^{-4}$.
The Mie scattering is denoted as $(1-f_{\Ray})(\frac{\lambda}{\lambda_0})^{b_{\Mie}}$.
$a_0$, $\lambda_{0}$, and $b_{\Mie}$ denote the same factors as above in \cref{eq.1}.
$f_{\Ray}$ and $1-f_{\Ray}$ indicate the fraction of Rayleigh scattering and Mie scattering, respectively.
The power factor 4 for Rayleigh scattering was determined in previous studies~\cite{Jacques2013OpticalReview}.
We have observed from our phantom experiments that the reduced scattering spectrum is affected by the cross-talk most noticeably at the wavelength of 400 - 450 nm and 1200 - 1700 nm where haemoglobin and water dominate the absorption spectrum (see \cref{fig:2}).
To that end, the $\mu'^{\IAD}_s (\cdot)$ is fitted with either of the above equations at $\lambda \in [460,\: 1150]\:nm$ to achieve optimum fitting results. We chose $\lambda_0 = 500\:nm$ because at this wavelength there is no significant absorption that arise from either haemoglobin or lipid or water.
For the non-linear least squares fitting, our implementation makes use of the \textrm{nlinfit} function in MATLAB\textsuperscript{\textregistered}.

For the 
\textbf{second-stage IAD}, at each wavelength $\lambda$, we constrain the scattering coefficient with the value from the model fitted in the first-stage IAD, $\mu'^{\RevIADMie}_s(\lambda)$ or $\mu'^{\RevIADMieRay}_s(\lambda)$.
The current version of the open-source IAD software (iad 3-9-12, 31 May 2017) already allows for providing a reduced scattering coefficient $\mu'_s(\lambda)$ as a constraint in the inverse calculation leading to the estimation of the absorption coefficient $\mu_a(\lambda)$.
However, it only allows so by inputting model parameters ($a_0$, $\lambda_0$ and $b_{\Mie}$) based on Mie theory.
%
To provide more flexibility and encompass both Mie and Rayleigh scattering,
we modified the IAD software to allow it to use model parameters ($a_0$, $\lambda_0$, $f_{\Ray}$ and $b_{\Mie}$) from \cref{eq.2} to calculate $\mu'_s(\lambda), \:\lambda \in [400,\:1800]\:nm$.
The modified source code is available at \url{https://github.com/cai4cai/RevIAD}.
This was deemed important to exploit the \emph{two-stage} IAD also in the case of biological samples with scatterings attributed to ultra-structures (e.g. collagen fibrils).
The final output of absorption and reduced scattering were denoted as $\mu^{\RevIADMie}_a$, $\mu^{\RevIADMieRay}_a$, and $\mu'^{\RevIADMie}_s$, $\mu'^{\RevIADMieRay}_s$, respectively. 

\subsection{Validation protocol with simulated data}
\begin{align} 
\eqalign{\mu_a^{\msim}(\lambda) &= [\mu_a^{\Hb}(\lambda) \hspace{5pt} \mu_a^{\HbOO}(\lambda) \hspace{5pt} \mu_a^{\Water}(\lambda) \hspace{5pt} \mu_a^{\Lipid}(\lambda)]
          \begin{bmatrix}
             \vspace{3pt}
          f^{\msim}_{\Hb}\\
          \vspace{3pt}
         f^{\msim}_{\HbOO}\\
         \vspace{3pt}
         f^{\msim}_{\Water}\\
         \vspace{3pt}
         f^{\msim}_{\Lipid}\\
         \end{bmatrix}}
         \label{eq.endmembers}
  \end{align}

\begin{figure}
	\centering
	\includegraphics[scale=0.13]{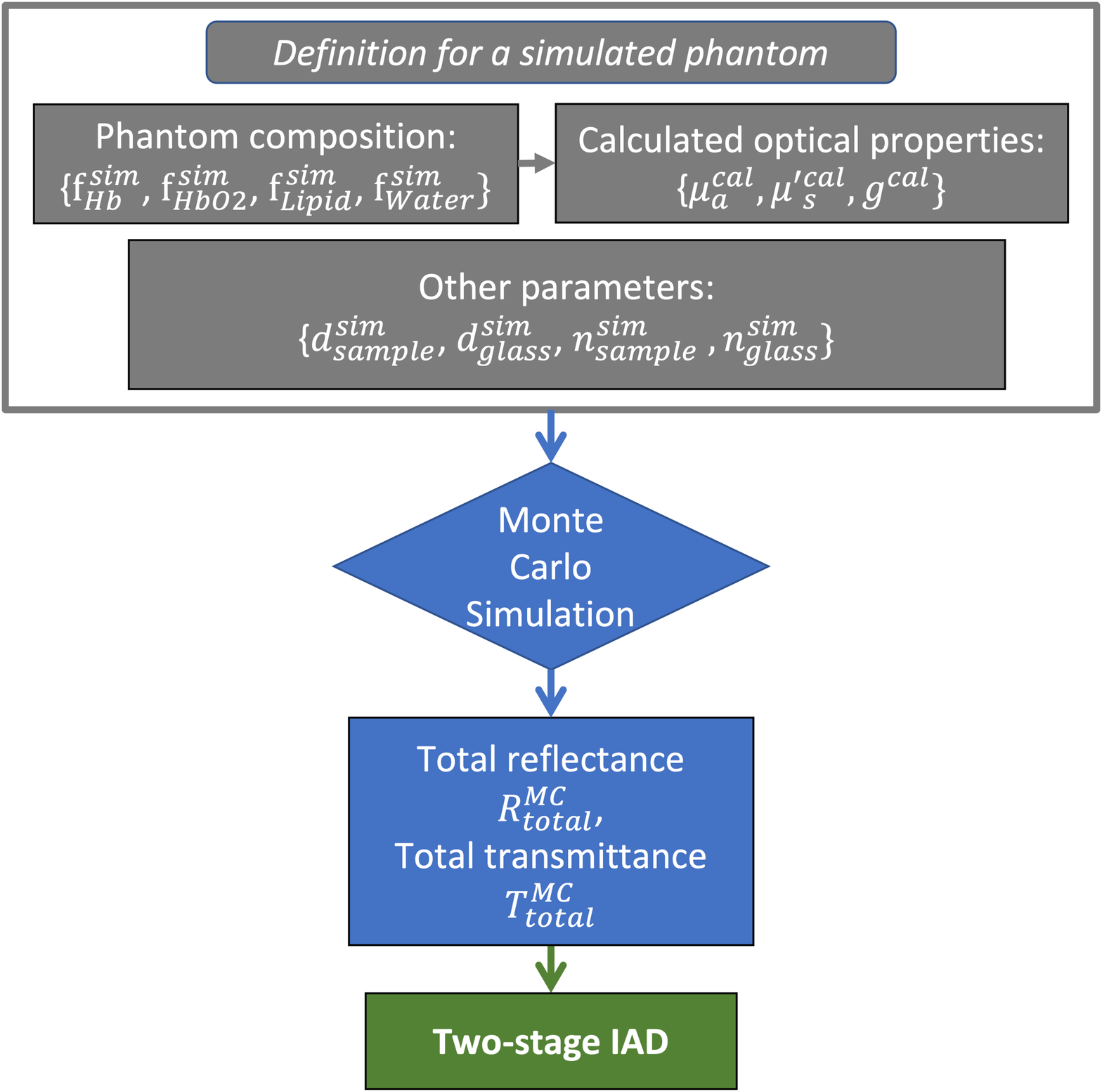}
	\caption{A flowchart diagram to illustrate the pipeline of the validation process with computer simulated data.}
	\label{fig:4}
\end{figure}

To validate the proposed two-stage IAD, we first used computer-synthetic phantoms and their transmittance and reflectance spectra which were simulated by Monte Carlo modeling of light transport in Multi-Layered tissues (MCML) algorithm\footnote{Available from \url{https://omlc.org/software/mc}} \cite{L.Wang1995MCML-MonteTissues,Wang1997ConvconvolutionTissues}. The work flow of the procedure includes the following steps:

\begin{enumerate}
\item Define a synthetic phantom consisting of four chromophores (Hb, $\HbOO$, Water, Lipid) that are the main constituents in most biological tissues. Assign each of them a volume fraction that are similar to those in biological tissues $f^{\msim}_{\Hb}, f^{\msim}_{\HbOO},f^{\msim}_{\Water}, f^{\msim}_{\Lipid}$ \cite{Nachabe2010EstimationNm}.
\item Calculate the total absorption spectrum $\mu^{\calculated}_a$ of the phantom based on the known respective absorption spectra of the four chromophores $\mu^{\Hb}_a,\mu^{\HbOO}_a, \mu^{\Water}_a, \mu^{\Lipid}_a$ in their pure form and their volume fractions as per \cref{eq.endmembers} \cite{Nachabe2010EstimationNm,Nachabe2010EstimationNmb};

We assume the reduced scattering $\mu'^{\calculated}_s$ and the anisotropy coefficient $g^{\calculated}$ are only attributed to the lipid content from the Intralipid. Thus we use the formula in \cref{eq.4} as presented in \cite{Aernouts2014DependentRange} to calculate the expected optical value of Intralipid ($\mu'^{\expected}_s$ and $g^{\expected}$) with given volume concentration.
\item Define the phantom’s thickness $d$, and refractive index $n$ to be used as input for the Monte Carlo simulation generating the total reflectance $R^{\simMC}_{\total}(\lambda)$ and total transmittance $T^{\simMC}_{\total}(\lambda)$ spectra of the phantom.
\item Use the IAD algorithm to calculate, at each wavelength $\lambda$, the absorption coefficient $\mu^{\simIAD}_a(\lambda)$ and reduced scattering coefficient $\mu'^{\simIAD}_s(\lambda)$ of the phantom from the $R^{\simMC}_{\total}(\lambda)$ and $T^{\simMC}_{\total}(\lambda)$.
\item Use nonlinear least squares to fit $\mu'^{\simIAD}_s(\cdot)$ with \cref{eq.1} and \cref{eq.2}.
\item In the second stage IAD, use the fitted reduced scattering $\mu'^{\simIADMie}_s(\lambda)$ or $\mu'^{\simIADMieRay}_s(\lambda)$as an input constraint for the IAD algorithm to calculate a revised absorption spectrum, and to obtain the revised absorption coefficient $\mu^{\simRevIAD}_a(\lambda)$.
\item For validation, compare the absorption and reduced scattering coefficients derived from the two-stage IAD (\(\mu^{\simRevIAD}_a(\lambda), \mu'^{\simRevIAD}_s(\lambda)\)) and those derived from the original IAD (\(\mu^{\simIAD}_a(\lambda), \mu'^{\simIAD}_s(\lambda)\)) to the ground-truth from the simulations (\(\mu^{\simIAD}_a(\lambda), \mu'^{\simIAD}_s(\lambda)\)).
\end{enumerate}

\subsection{Validation protocol with integrating sphere measurements on physical phantoms}
\label{sec:phantom}
\begin{figure}
	\centering
	\includegraphics[scale=0.13]{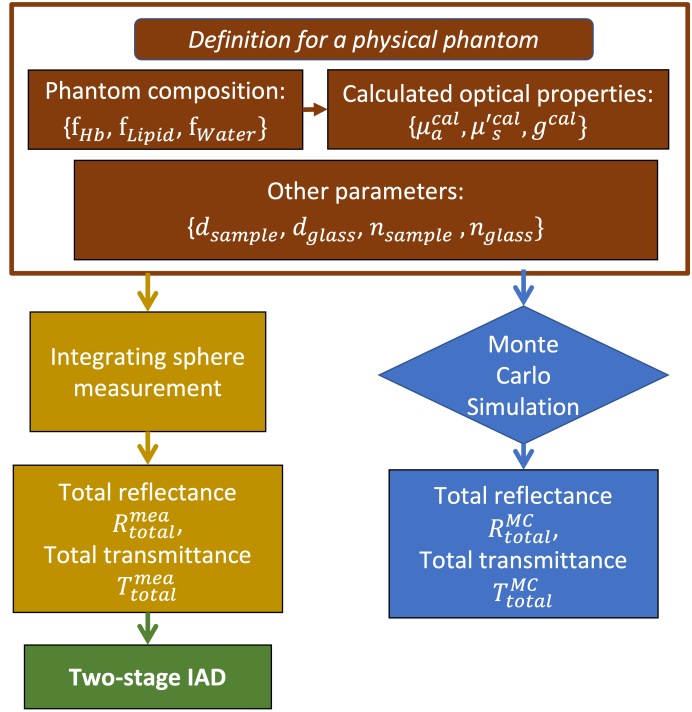}
	\caption{A flowchart diagram to illustrate the validation process with integrating sphere measurements.}
	\label{fig:5}
\end{figure}

We further validated our method using standard physical phantoms of which the absorption coefficient and reduced scattering coefficient can be precisely calculated from the phantoms' compositions.
The procedure was similar to the one for validation on simulated data, except that we used the spectrophotometer to measure the total transmittance and total reflectance of the physical phantom (workflow shown in \cref{fig:5}).

\subsubsection{Phantom preparation.}
We prepared liquid phantoms using lipid emulsion (Intralipid\textsuperscript{\textregistered} 20\%, batch numbers 10PE3391 EXP:04/2022, Fresenius Kabi Ltd, UK) as a scattering agent and haemoglobin (lyophilized powder, H2500 Sigma-Aldrich, UK) as absorber.
Intralipid has been demonstrated as a reliable material for introducing controllable scattering properties in phantoms.
Its optical properties and stability has also been fully investigated and reported (\cite{Michels2008OpticalEmulsions,DiNinni2012FatPhantoms,Aernouts2014DependentRange,Spinelli2014DeterminationInk,Krauter2015OpticalParameters,Samkoe2017DevelopmentExcision,Fredriksson2017EvaluationPhantoms}).
The batch-to-batch variations of the reduced scattering coefficient of Intralipid were found to be about 1.5\% \cite{DiNinni2011EffectPhantoms}.
Its predominant absorbers are soy bean oil and water. 
Thus its absorption coefficient is negligible in the visible wavelength range and is practically equal to that of water at the near infrared range \cite{Michels2008OpticalEmulsions}.
Haemoglobin solution has been commonly used in optical phantom preparation to mimic blood absorption spectrum \cite{Haj-Hosseini2014DevelopmentPhantom,James2014IntegratedPhantom,Ghassemi2015RapidImaging,Rejmstad2017ASpectroscopy}.
While its extinction coefficient has been determined and is available from literature \cite{Pogue2006ReviewDosimetry,Lamouche2012,Haj-Hosseini2014DevelopmentPhantom,DuLe2014MeasurementsMicrospheres}, 
it should be noted that the lyophilized powder of haemoglobin used in this study may be predominantly methaemoglobin (MetHb), since the native haemoglobin is readily oxidized in air during the supplier's manufacturing process.
The extinction coefficient of MetHb is distinctly different to either oxygenated haemoglobin (HbO) or deoxygenated haemoglobin (Hb) \cite{Cruz-Landeira2002DeterminationSpectrophotometry,Zijlstra1997SpectrophotometryMethemoglobin}.
We thus used our spectrophotometer to measure the absorption coefficient 
($\mu_a^{\MetHb}$)
of the methaemoglobin/haemoglobin solution prepared from lyophilized powder of bovine haemoglobin and Phosphate Buffered Saline (PBS, Sigma Alrich, UK) at concentration of 1.0 \( mg/ml\), and used as an unit absorption coefficient in later calculations.
We also calculated the absorption coefficient of 10\% Intralipid (IL) as reported in the work of Michels et al.~\cite{Michels2008OpticalEmulsions}.
In summary, the basic absorption coefficient spectra of the substances (\( \mu_a^{\IL10}, \mu_a^{\MetHb}, \mu_a^{\Water}\)) we used for the phantom preparation are shown in SI Figure 1.

We prepared six sets of phantoms with various combinations of IL and MetHb concentrations  ($C_{\IL}$ and $C_{\MetHb}$ ) as shown in \cref{tab.2}. All phantoms were prepared with PBS at room temperature (ca. $21^{\circ}\text{C}$) and were mixed with a vortex mixer (Fisherbrand, Fisher Scientific UK) for 2 min.
We chose PBS for preparing phantoms to stabilise the mixture of methaemoglobin, lipid emulsion and water, and to ensure the reproducibility of the measurements.
Intralipid is a colloidal suspension stabilised by surface positive charges on the colloidal particles creating an electrical double layer according to the Derjaguin-Landau-Verwey-Overbeek (DVLO) theory\cite{Horinek2014}. Methaemoglobin is able to reversibly bind to H+ ions therefore increasing the pH of the deionised water and the availability of negatively charged ions.
When mixed with intralipid, these negative ions neutralise the surface charges on the colloids removing the electrical double layer and cause the suspension to aggregate and therefore the Methaemoglobin and Intralipid solutions are immiscible.
By introducing PBS, the pH of both solutions is maintained, thus preventing the neutralisation of these surface charges and therefore preventing the aggregation of the intralipid colloids.
For this reason, PBS is included in all phantom synthesis to ensure an even dispersion of intralipid by maintaining a constant pH.

\begin{table}
\caption{\label{tab.2}Phantom compositions and measurement information}
\footnotesize
\begin{tabular}{p{3cm}|p{1.5cm}p{1.5cm}p{1.5cm}p{1.5cm}p{1.5cm}p{1.5cm}}
\br
&Set A&Set B&Set C&Set D&Set E&Set F\\
\hline
$C_{\MetHb}$ (mg/ml)&2.5&5.0&7.5&10.0&1.5&3.0\\
$C_{\IL}$ (v/v \% )&2.0&2.0&2.0&2.0&1.0&6.0\\
Total volume (ml)&10&10&10&10&10&10\\
IS port size &Original &Original&Original&Original&Original, modified&Original, modified\\
\br
\end{tabular}\\
IS: Integrating sphere
\end{table}

The expected absorption and scattering coefficient of the liquid phantom were calculated as follows:
\begin{equation}
   \mu_a^{\expected}(\lambda) = \mu_a^{\MetHb}(\lambda)\times C_{\MetHb} + \mu_a^{\IL10}(\lambda) \times C_{\IL}  + \mu_a^{\Water}(\lambda) \times (1 - C_{\IL}), \: \lambda \in [400,\: 1800]\:nm
  \label{eq.3}
\end{equation}
\begin{equation}
\begin{split}
\mu_s(\lambda) &= \frac{1.868\mathrm{e}{-10}\lambda^{-2.59}}{0.227}\Phi_{\IL}\frac{(1-\Phi_{\IL})^{p(\lambda)+1}}{(1+\Phi_{\IL}(p(\lambda)-1))^{p(\lambda)-1}}\\
p(\lambda) &= 1.31 + 0.0005481 \lambda\\
g(\lambda) &= 1.1- (0.58\mathrm{e}{-3} \lambda)\\
\mu_s^{'\expected}(\lambda) &= \mu_s(\lambda) g(\lambda), \: \lambda \in [400,\: 1800]\:nm \\
\end{split}
\label{eq.4}
\end{equation}
where $\Phi_{\IL}$ is the volume concentration of Intralipid. The refractive index of the phantom was calculated based on the volume concentration of scattering particles or lipid ($\Phi_{\IL}$) in the phantom $n_{\textrm{phantom}} = n_{\Water} + 0.14\Phi_{\IL}$ \cite{Aernouts2013}.
We also designed and developed sample chambers to contain liquid phantom samples for integrating sphere measurements. The sample chamber consisted of two glass slides 52 mm (w) x 76 mm (h) x 1 mm (d), and a 2 mm thick U shape acrylic spacer that was processed by a laser cutter (VLS6.75, Universal Laser Systems, USA) and glued between the two glass slides with a UV-curing optical adhesive (NOA68, Thorlabs, USA).

\subsubsection{Small sample kit.}
To validate the measurement with a small sample kit, we used the liquid phantoms developed as described in 
\cref{sec:phantom}.
The phantoms' total transmittance and total reflectance were measured in the spectrophotometer with the 3 mm and 5 mm sample kit.

\subsubsection{Measures of accuracy.}
We used the median value of residuals, Mean Squared Error (MSE), and Pearson Correlation Coefficient (CC) to characterise the performance of the two-stage IAD algorithm as compared with the original version.
MSE was calculated by $\textrm{MSE} = \dfrac{1}{n}\sum_{i=1}^{n}(y_i^{\textrm{RevIAD}} - y_i^{\IAD})$, and CC was calculated using MATLAB\textsuperscript{\textregistered} \textit{corrcoef} function. 

\subsection{Measurement on fresh mouse tissue} 
To evaluate the capability and performance of the two-stage IAD algorithm and the small sample kit on measuring biological tissue, we performed \textit{ex vivo} testing on various mouse tissues. All experiments were performed in accordance with local laboratory guidelines. A female laboratory Balb/C mouse weighing of approximately 30 g was used in the experiments. Animal was culled by rising the concentration of $\textrm{CO}_2$, followed by a secondary confirmation with cervical dislocation. Immediately after dissection, the tissue specimens were rinsed with PBS solution for about 1 min to remove any excess blood on their surface. One specimen of liver, one specimen of quadriceps muscle, and one specimen of brain (cerebral cortex) were examined and reported here. The average size of specimens was 7 x 10 x 2 $\textrm{mm}^3$. The specimens were placed in the sample chamber between two glass slides. We used index matching gel (G608N3, Thorlabs, UK) to secure the tissue in place by filling the empty region within the specimen chamber at either end of the tissue and to diminish refractive index mismatch at the tissue/glass boundaries. The specimens were brought for spectrophotometer measurements promptly (within 15 min).

\section{Results \& discussions}
\subsection{Validation of the two-stage IAD algorithm}
\subsubsection{Computer simulations.}
\begin{figure}
	\centering
	\includegraphics[scale=0.3]{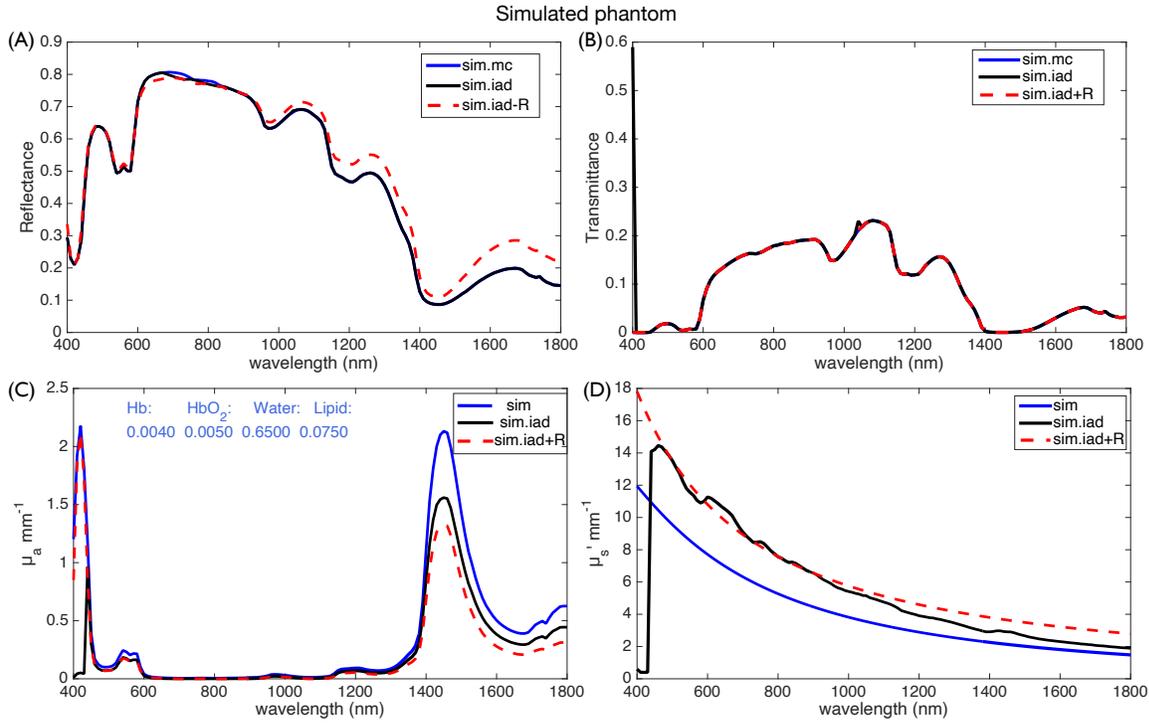}
	\caption{Data of a computer simulated phantom which comprises oxy- and deoxy- haemoglobin (HbO2, Hb), water and lipid with given volume fractions. A-B: Total reflectance and total transmittance spectra of the synthetic phantom A, generated by Monte Carlo simulation, and IAD respectively. C-D: Absorption and reduced scattering spectra of simulated phantom that are calculated with the two-stage IAD respectively. It is worth noting that in the longer wavelength range ($>$1400 nm) the proposed two-stage IAD method deteriorated the reconstruction of absorption spectrum \(\mu_a^{\RevIAD}\) (C). Presumably, this was because the simple power law fitting with input constraints used in the second stage IAD calculation was not capable of accurately modelling the reduced scattering spectrum over a broad range of wavelength (D).}
	\label{fig:7}
\end{figure}

The computer simulated phantom comprised oxy- and deoxy- haemoglobin (HbO, Hb), water and lipid with given volume fractions as displayed in \cref{fig:7}.
Results of absorption coefficient spectra and reduced scattering spectra ($\mu_a^{\simRevIAD}, \mu_s'^{\simRevIAD}$) of the phantom using the proposed two-stage IAD process are presented in \cref{fig:7}.
The values are compared to results by original IAD algorithm ($\mu_a^{\simIAD}, \mu_s'^{\simIAD}$) and as well as the calculated expected values ($\mu_a^{\msim}, \mu_s'^{\msim}$).
There is a noticeable discrepancy between the IAD derived $\mu_a^{\simIAD}$ and the calculated expected \(\mu_a^{\msim}\) at \(\lambda = 420\: nm\) which is affected by the cross-talk with \(\mu_s'^{\simIAD}\) as shown in \cref{fig:7}(C, D).
The underestimated absorption is successfully corrected by applying the second stage of IAD as proposed.
The corrected $\mu_a^{\simRevIAD}$ aligns greatly with the expected \(\mu_a^{\msim}\) (\cref{fig:7}(C, D)). 
However, we noticed that in the longer wavelength region when \(\lambda \geq 1400\: nm\) there is noticeable discrepancy between the resulting values and the expected value.
This is presumably caused by using the simple power law fitted \(\mu_s^{\IAD}\) as the constrain for the second stage of IAD computing, the residuals of the fitting model are higher at the longer wavelength range. Simulated transmittance spectra and reflectance spectra by using MCML algorithm and the values were derived with IAD algorithms are also presented in \cref{fig:7}(A, B). 
As a result of two-stage IAD the pronounced discrepancy of the transmittance spectra at \(\lambda = 420\: nm\) was rectified.
We also notice that $\mu_a^{\simRevIAD}, \mu_s'^{\simRevIAD}$ mostly changed the reflectance spectra rather than transmission spectra.

\begin{figure}[htb!]
	\centering
	\includegraphics[scale=0.80]{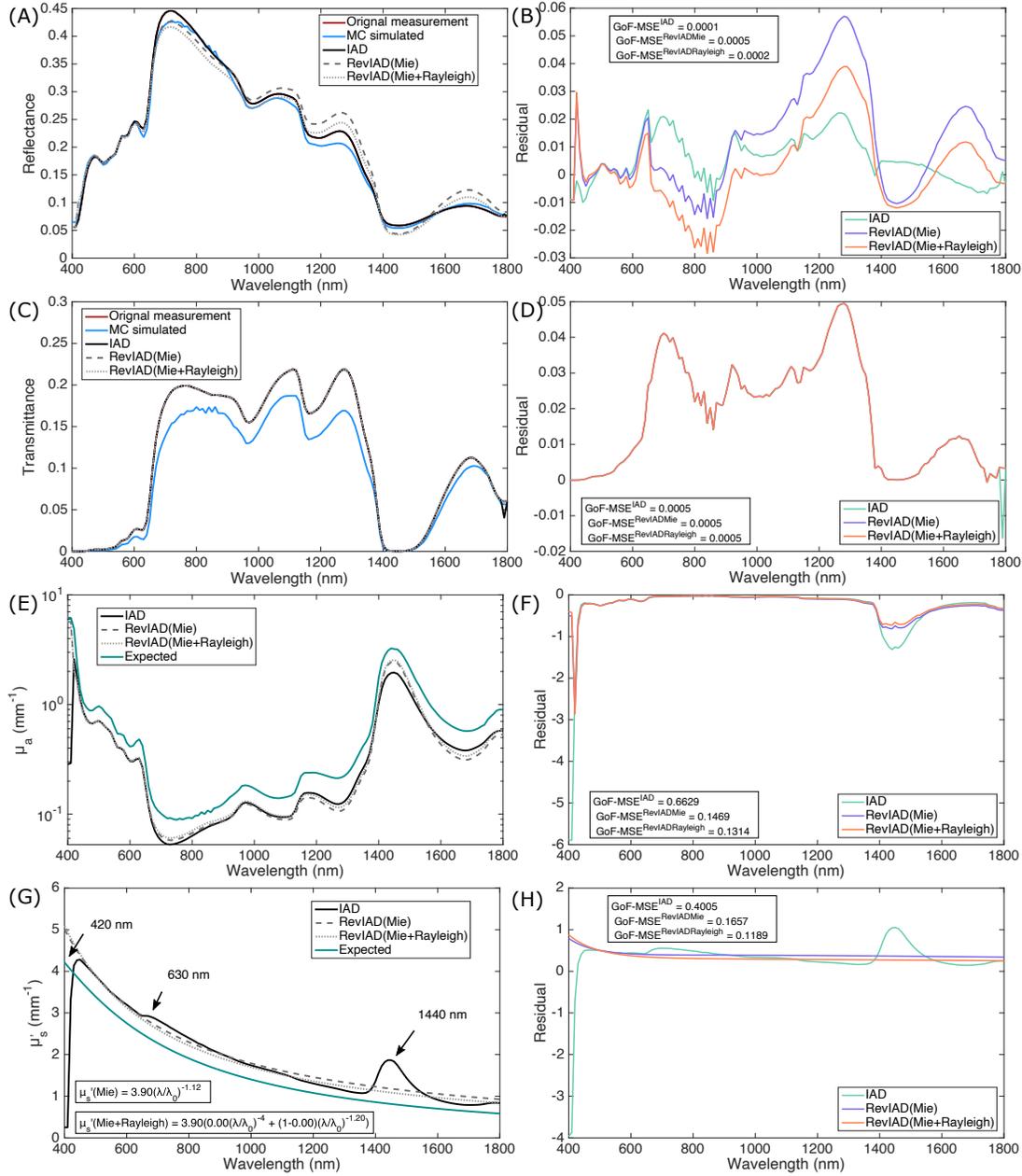}
	\caption{A representative example (Phantom set C) of phantom experiments and results of the two-stage IAD algorithm.}
	\label{fig:8}
\end{figure}

\subsubsection{Phantom experiments}
Phantom set A-F were freshly produced and measured with the integrating sphere using the standard set-up. The phantom sample dimension was 40 mm (w) x 50 mm (h) x 2 mm (d) which was defined by the size of the sample chamber.
A representative example (phantom set C) of the integration sphere measurements and calculated optical properties is shown in \cref{fig:8}.
We compared the transmittance spectra and reflectance spectra calculated by using the IAD algorithm to the values from integrating sphere measurements and simulated values by using MCML algorithm (\cref{fig:8}(A, C), residuals are plotted in \cref{fig:8}(B, D).
The absorption coefficient spectra and reduced scattering spectra ($\mu_a^{\RevIADMie}$, $\mu_s'^{\RevIADMie}$; $\mu_a^{\RevIADMieRay}$, $\mu_s'^{\RevIADMieRay}$) of phantom C are presented in \cref{fig:8}(E, G), and residuals as compared with expected values are plotted in \cref{fig:8}(F, H).
For the absorption coefficient, the results from the two-stage IAD are compared to values obtained with original IAD algorithm ($\mu_a^{\IAD}$) and as well as to the calculated expected values ($\mu_a^{\expected}$), showing an improved mean MSE (averaged over all six phantom sets) from 0.6629 to 0.1469 and 0.1189 when using the Mie fitted scattering and Mie+Rayleigh fitted scattering respectively.
There is an constant offset across whole spectrum between the \(\mu_s'^{RevIAD}\) and the calculated expected values ($\mu_s'^{\expected}$), which is likely due to 1) the discrepancy between Intralipid batches; 2) the variability of different models of estimating intralipid's reduced scattering (SI Figure 2); 3) the model we used to create the expected reduced scattering values is given at the 600-1850 nm range. In addition, the Mie theory based fitting was done at $\lambda \in [460,\: 1150]\:nm$ to eliminate outliers caused by cross-talk, future works on using robust fitting on the whole wavelength range would reduce the discrepancy.
Nevertheless, the peaks caused by cross-talk (420 nm, 630 nm and 1440 nm) in the reduced scattering spectrum were removed (as marked in \cref{fig:8} with arrows.).
We calculated the mean squared error, median value of the residual and Person's correlation coefficient of each phantom experiment, their mean values averaged from six phantom sets are demonstrated in \cref{tab.IADvsExpected}.
The difference between the IAD calculated values and the expected values are reduced markedly for absorption and reduced scattering as reflected in the reduced MSE value, while the similarity between them are increased to yield a higher correlation score.

\begin{table}[htb!]
\caption{\label{tab.IADvsExpected}Measures of discrepancy between IAD results and the expected values}
{\scriptsize
\begin{tabular}{m{2.3cm}|m{0.7cm}m{0.7cm}m{0.7cm}m{0.9cm}|m{1.2cm}m{0.7cm}m{0.7cm}m{1.2cm}|m{0.7cm}m{0.7cm}m{0.7cm}m{0.7cm}}
\br
&\multicolumn{4}{c}{MSE}&\multicolumn{4}{|c}{Median}& \multicolumn{4}{|c}{CC}\\
\cline{2-13}
&R&T&$\mu_a$&$\mu'_s$&R&T&$\mu_a$&$\mu'_s$&R&T&$\mu_a$&$\mu'_s$\\
 \hline
IAD &0.0008&0.0024&0.4977&0.4140&$-0.0093$&0.0076&0.3831&$-0.0732$&0.9955&0.9271&0.8739&0.8743\\
RevIAD (Mie)&0.0011&0.0008&0.3454&0.1359&$-0.0036$&0.0076&0.4309&$-0.0831$&0.9871&0.9929&0.9991&0.9685\\
RevIAD (Mie+Rayleigh)&0.0013&0.0008&0.2955&0.1246&$-0.0074$&0.0076&0.3568&$-0.0777$&0.9887&0.9929&0.9987&0.9709\\
\br
\end{tabular}}\\
{\small MSE: Mean Squared Error, CC: Person Correlation Coefficient}
\end{table}

\begin{figure}
	\centering
	\includegraphics[scale=0.8]{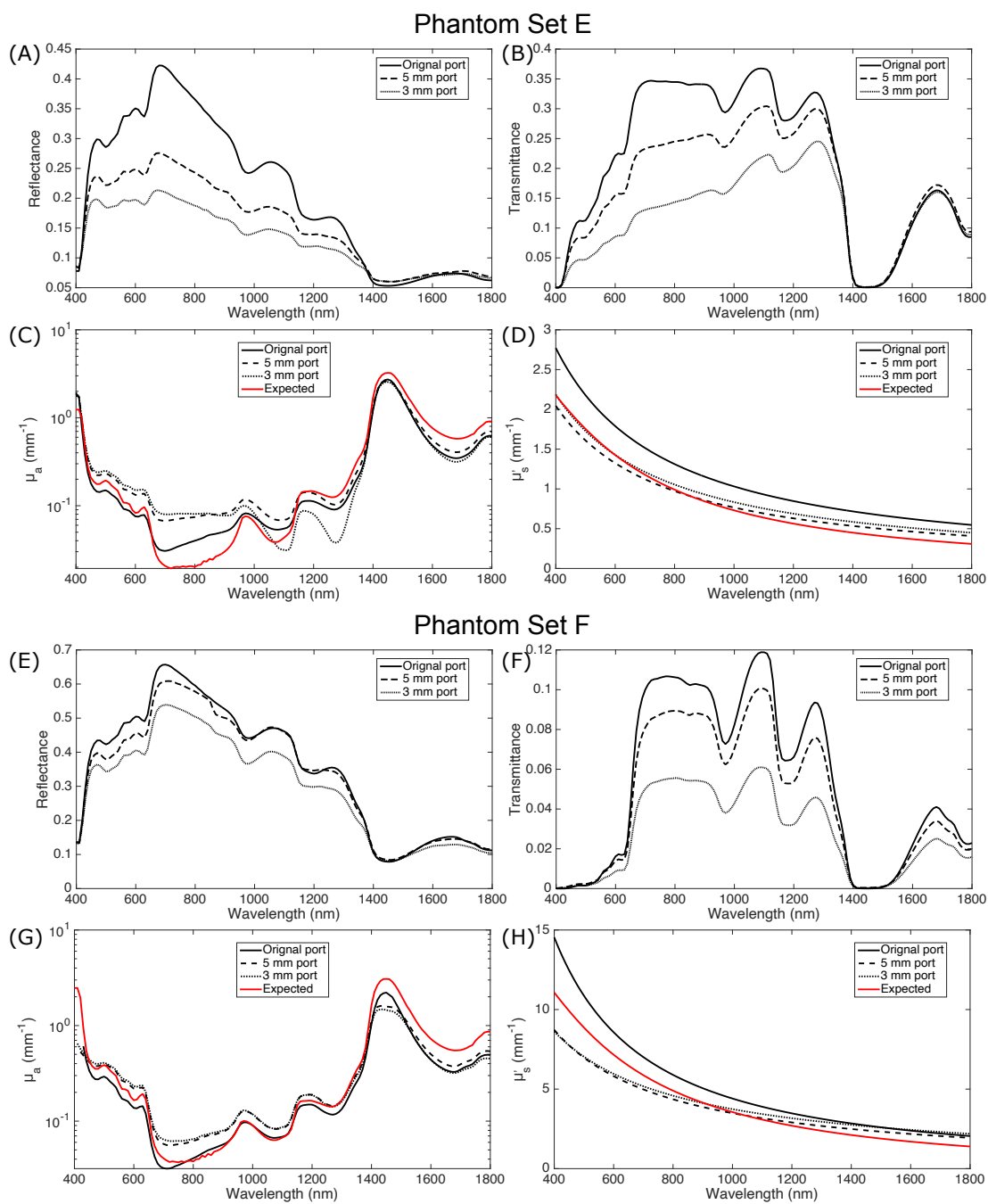}
	\caption{Spectrophotometer integrating sphere measurements (A, B) of small samples of liquid phantom using the small sample kit and the converted optical properties using the two-stage IAD as validated against the original sample port. Optical absorption coefficients (C) and Reduced scattering (D) of liquid phantom using different port sizes.}
	\label{fig:9}
\end{figure}

\subsection{Validation of the small sample kit method}
The liquid phantom set B, E, F from the same freshly-prepared batch were then measured with the small sample kit.
The phantom sample dimension was 7 mm (w) x 15 mm (h) x 2 mm (d) defined by the size of the sample chamber. As expected, the reflectance and transmittance spectra measured with small sample kits were lower than the ones with original set-ups, which were attributed to the light loss on the lateral sides of the sample.
The distance (\(h_{t}\)) from the edge of the incident beam on the sample to the edge of the adapted integrating sphere transmittance ports were calculated as 1 mm for the 3 mm port and 1.5 mm for the 5 mm port.
\(h_r\) for the reflectance ports were calculated as 0.5 mm and 0.75 mm for the 3 mm port and 5 mm port, respectively.
For the liquid phantom sample we used here, the \(1/(\mu_a + \mu_s')\) value is greater than \(1/8\) roughly at wavelength range \(\lambda \in [430 \: 1400] \: nm\). Consequently, the \(h\) value is less than fives of \(1/(\mu_a + \mu_s')\) resulting in noticeable light loss \cite{Mesradi2013OpticalRat}.
As shown in \cref{fig:9}, residuals in reflectance and transmittance measurements are noticeably larger than residuals in the calculated optical properties, suggesting that the IAD algorithm with MC as subroutine is able to consider the sample port size and incident light beam size and eliminate their influence on the resulting optical properties.
However, as compared with measurement with original port, \(\mu_a\) and \(\mu_s'\) obtained with either small size port still display an offset across entire spectra. This is likely because the lost light from the lateral sides of the sample and as well as on the sides of glass slides for sandwiching the sample can not be fully recovered by IAD. Vinvely et al. studied the lost light during IAD calculation in relation to the sample size (optical thickness, radial path length), beam size and sample port size \cite{10.1117/12.2546265}. They found when the incident beam had a diameter less than 80\% of the sample length, the optical properties of the sample could be well estimated. The work from Prahl also demonstrated that the inherent limitation of IAD \cite{Prahl2011EverythingAdding-doubling}. Our results are consistent with their findings that the inherent challenging of using the IAD on small samples can be addressed to some extent, and a quantified correction model to mitigate the light loss effects could be future directions.


\subsection{Optical properties of fresh mouse tissues}
\begin{figure}[h!]
	\centering
	\includegraphics[scale=0.8]{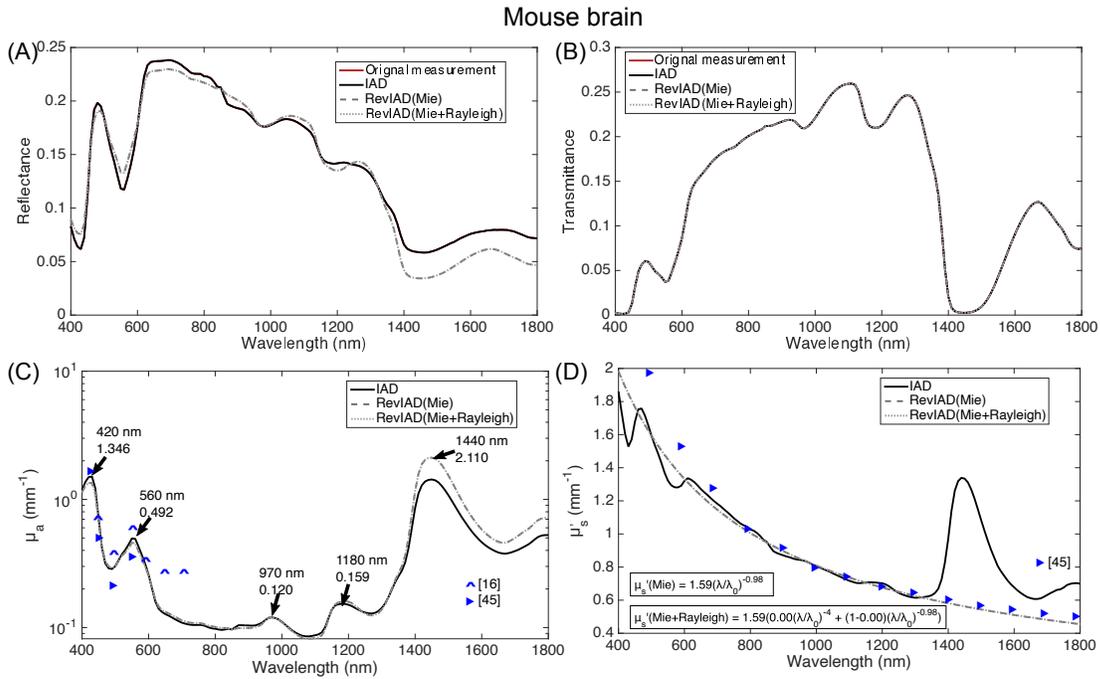}
	\caption{Reflectance (A) and transmittance (B) spectra, absorption coefficient (C) and reduced scattering coefficient spectra (D) of freshly dissected mouse brain tissue.}
	\label{fig:12}
\end{figure}

\begin{figure}[h!]
	\centering
	\includegraphics[scale=0.8]{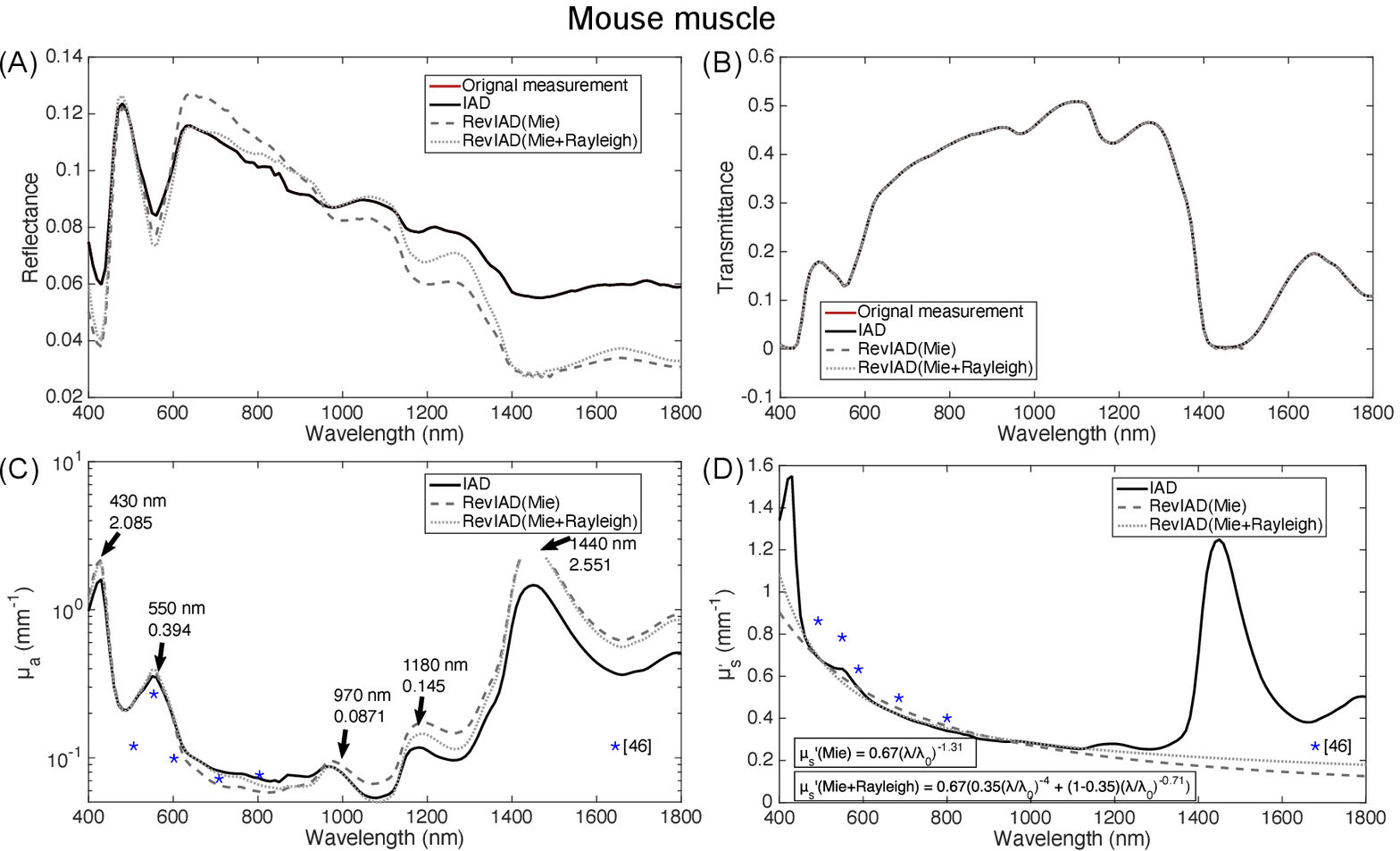}
	\caption{Reflectance (A) and transmittance (B) spectra, absorption coefficient (C) and reduced scattering coefficient spectra (D) of freshly dissected mouse muscle tissue.}
	\label{fig:13}
\end{figure}

\begin{figure}[h!]
	\centering
	\includegraphics[scale=0.8]{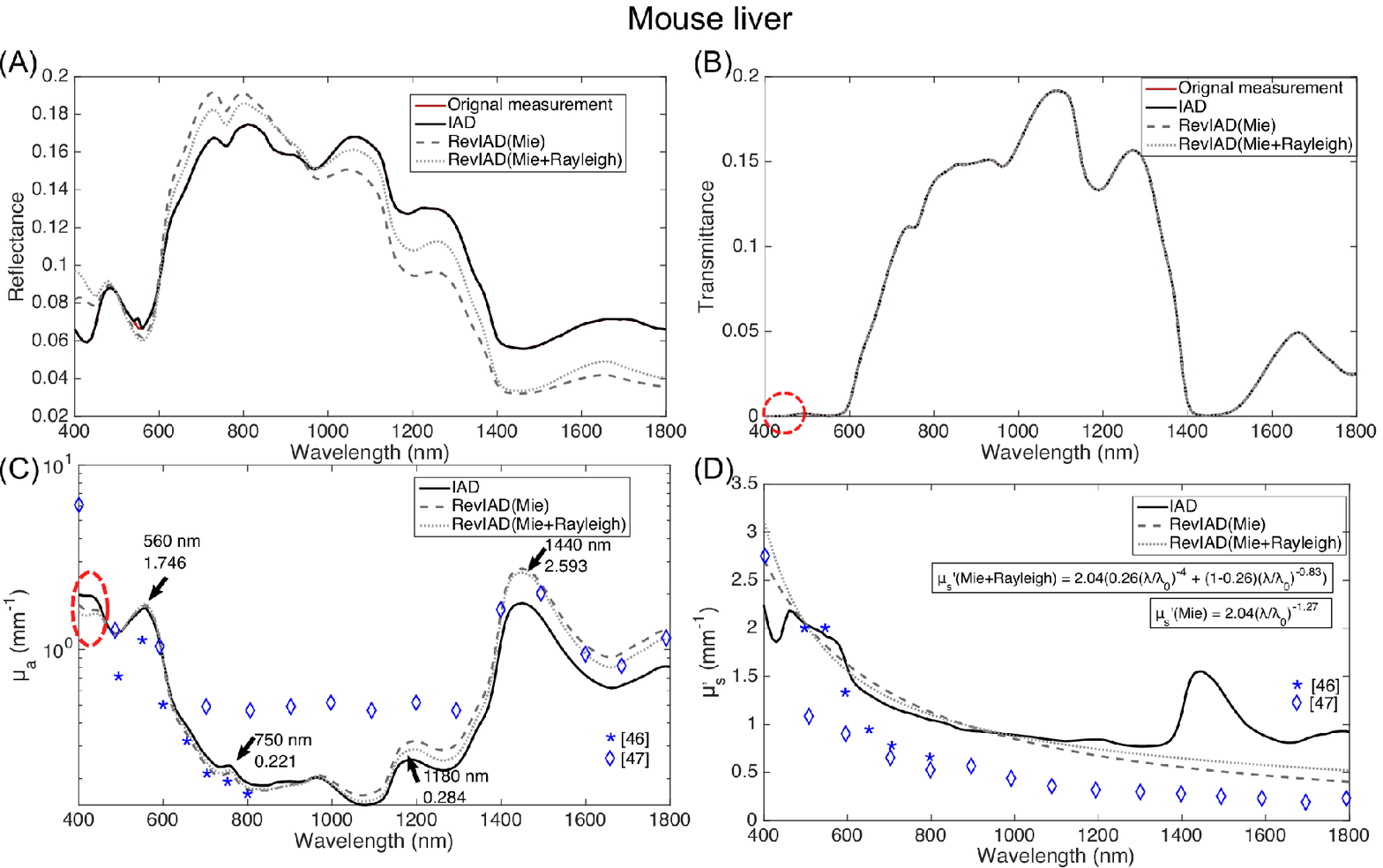}
	\caption{Reflectance (A) and transmittance (B) spectra, absorption coefficient (C) and reduced scattering coefficient spectra (D) of freshly dissected mouse liver tissue.}
	\label{fig:14}
\end{figure}

Tissue specimens were measured with a spectrophotometer using our small sample kit: 3 mm port for the brain tissue and 5 mm port for the others.
For IAD calculation the refractive index of brain tissue was selected at 1.36 \cite{Binding2011BrainMicroscopy}, and the n value of muscle and liver tissue were selected at 1.37 \cite{Giannios2016VisibleMalignancies,Tuchina2015QuantificationMyocardium}. The anisotropy value g was fixed at 0.85 for all tissue types, this was chosen based on our calculation (SI Figure 3) as well as the reported values in previous works \cite{Gebhart2006InAdding-doubling,Honda2018DeterminationAcid,Tuchin2007OpticalScattering}.  
As shown in \cref{fig:12}, \cref{fig:13} and \cref{fig:14}, the water absorption with maxima at 970, 1180, and 1440 nm are evident in the all tissue specimen. The total haemoglobin absorption bands with maxima at 420/430 and 550 nm present in the brain and muscle absorption spectrum with a lower magnitude as compared with the liver spectrum, which is because of the higher blood content in the liver tissue. However, liver's absorption value at 420/430 nm is greatly lower than expected (as highlighted in red dash circle in \cref{fig:14}) which is presumably caused by extremely low transmittance ($\sim$0) and thus can't be accurately processed by IAD. Also, there is a noticeable peak in the liver's absorption spectrum at 750 nm, corresponding to the de-oxygenated hemoglobin's absorption peak at around 758 nm. This observation suggests that the animal's blood oxygenation level was low when it was sacrificed, which coincides with the termination method of using $\textrm{CO}_2$. The reduced scattering coefficient decreases as wavelength increases and the spectra can be fitted with Mie and Rayleigh theory as discussed in \cref{sec.reviad}.

We compared our results with data from related reported works in which fresh rat tissues have been investigated \cite{Mesradi2013OpticalRat,Golovynskyi2018OpticalApplicationsb,Nilsson1995MeasurementsTherapy,Parsa1989OpticalNm}, and these are included and plotted in \cref{fig:12}, \cref{fig:13} and \cref{fig:14} with respective symbols. For all tissue, the absorption coefficients and reduced scattering coefficients in the present work are generally in good agreement with the reported values. We also noted there are discrepancies in optical properties of liver tissue. The differences are within one order of magnitude which could possibly be caused by tissue variability and preparation variance.

\section{Conclusions}
The contributions of the present study are two folds: 1) the proposed two-stage IAD process is able to produce a revised absorption spectrum using the fitted reduced scattering as an input constraint at the second stage IAD; 2) the small sample kit allows characterisation of optical properties of small sized biological tissues \textit{ex vivo}.

The revised absorption spectra as well as the fitted reduced scattering spectra were evaluated with both computer simulations and physical phantom experiments. Results were compared with expected values that were calculated using well-established equations.
In general, the findings in the quantitative measures (Median of residuals, Mean Squared Error, and Pearson Correlation Coefficient) suggested that the revised absorption coefficients and fitted reduced scattering coefficients were closer to the expected values as compared to values obtained by original IAD algorithm.
In regard to the small sample kit, the incident beam was condensed as small as possible given the limited space in the spectrophotometer's sample chamber for modifying the optics. However, the results of using small sample kit, notably the absorption coefficients were higher (approx.
0.1 - 0.2 $\textrm{mm}^{-1}$) than the ones obtained using standard port.
Even though, our measurements of animal tissue were comparable to those previously reported values.
Future works would focus on provide a computer model that can simulate relations between sample port sizes and estimated optical properties to quantify the discrepancies.

Nevertheless,our proposed method of a two-stage IAD pipeline and small sample kit could be a useful and reliable tool to characterise optical properties of biological tissue \textit{ex vivo} particularly when only small size samples are available. Our method and the associated data represent a valuable resource, which is being made publicly available, that can be used as comparison with other related works. 

\section{Disclosure}
ME, JS, and TV are shareholders of Hypervision Surgical Ltd London, UK, and have an equity interest in the company. TV is a shareholder of Mauna Kea Technologies, Paris, France.

\ack
This work was supported by the Wellcome Trust [203145Z/16/Z; 203148/Z/16/Z; WT106882], EPSRC [NS/A000050/1; NS/A000049/1], and by core funding from the Wellcome/EPSRC Centre for Medical Engineering [WT203148/Z/16/Z]. YX is supported by the L’Oréal-UNESCO UK and Ireland For Women in Science Rising Talent Programme. TV is supported by a Medtronic / Royal Academy of Engineering Research Chair [RCSRF1819\textbackslash7\textbackslash34]. ME is supported by the Royal Academy of Engineering under the Enter-prise Fellowship Scheme [EF2021\textbackslash10\textbackslash110]. I would like to thank Jana Kim for providing animal tissues and technical insight on animal tissue preparations. 
\clearpage
\section{Reference}
\bibliographystyle{iopart-num}
\bibliography{Methodreferences}

\end{document}